\documentclass[a4paper]{aa}
\usepackage{txfonts,graphicx}
\newcommand{\fo}{\ensuremath{f^\mathrm{o}}}
\newcommand{\fe}{\ensuremath{f^\mathrm{e}}}

\newcommand{\Den}{\ensuremath{f^\mathrm{o} + f^\mathrm{e}}}
\newcommand{\alphasun}{\ensuremath{\alpha_\odot}}
\newcommand{\deltasun}{\ensuremath{\delta_\odot}}
\newcommand{\alphaobj}{\ensuremath{\alpha_\mathrm{T}}}
\newcommand{\deltaobj}{\ensuremath{\delta_\mathrm{T}}}
\newcommand{\pq}{\ensuremath{P_Q}}
\newcommand{\pu}{\ensuremath{P_U}}
\newcommand{\Thetar}{\ensuremath{\Theta'}}
\begin{document}
\title{Exploring the surface properties of Transneptunian Objects and
       Centaurs with polarimetric FORS1/VLT observations
       \thanks{Based on observations made with ESO Telescopes at the 
               Paranal Observatory under programme ID 69.C-0133
               and 073.C-0561 (PI: H.\ Boehnhardt)}}
       \author{
        S.~Bagnulo      \inst{1}
       \and
        H.~Boehnhardt   \inst{2}
       \and
        K.~Muinonen     \inst{3}
       \and
        L.~Kolokolova   \inst{4}
       \and
        I.~Belskaya     \inst{5}
       \and
        M.A.~Barucci    \inst{6}
        }
\institute{European Southern Observatory,
           Alonso de Cordova 3107, Vitacura,
           Santiago, Chile.
           \email{sbagnulo@eso.org}
           \and
           Max-Planck-Institut f\"{u}r Sonnensystemforschung,
           Max-Planck-Strasse 2,
           37191 Katlenburg-Lindau,
           Germany.\\
           \email{hboehnha@linmpi.mpg.de}
           \and
           Observatory, PO Box 14, 00014 University of Helsinki, Finland.
           \email{muinonen@cc.helsinki.fi}
           \and
           University of Maryland, College Park, MD, USA.
           \email{ludmilla@astro.umd.edu}
           \and
           Astronomical observatory of Kharkiv National University, 
           35 Sumska str., 61022 Kharkiv, Ukraine.
           \email{irina@astron.kharkov.ua}
           \and
           LESIA, Observatoire de Paris, 5,
           pl.~Jules Janssen, FR-92195 Meudon cedex, France.
           \email{antonella.barucci@obspm.fr}
           }
\date{Received: 15 November 2005 / Accepted: 3 January 2006}

\abstract{
Polarization is a powerful remote-sensing method to investigate solar
system bodies. It is an especially sensitive diagnostic tool to reveal
physical properties of the bodies whose observational characteristics
are governed by small scatterers (dust, regolith surfaces). For these
objects, at small phase angles, a negative polarization is observed,
i.e., the electric vector $\vec{E}$ oscillates predominantly in the
scattering plane, contrary to what is typical for rather smooth
homogeneous surfaces.  The behavior of negative polarization with
phase angle depends on the size, composition and packing of the
scatterers. These characteristics can be unveiled by modelling the
light scattering by the dust or regolith in terms of the coherent
backscattering mechanism.
}
{We investigate the surface properties of TNOs and Centaurs by means
of polarimetric observations with FORS1 of the ESO VLT.
}  
{
We have obtained new broadband polarimetric measurements over
a range of phase angles for a TNO, 50000\,Quaoar (in the $R$
Bessel filter), and a Centaur, 2060\,Chiron (in the $BVR$
Bessel filters).  Simultaneously to the polarimetry, we have obtained
$R$ broadband photometry for both objects. We have modelled these new
observations of Quaoar and Chiron, and revised the modelling of
previous observations of the TNO 28978\,Ixion using an
improved value of its geometric albedo.
}
{
TNOs Ixion and Quaoar, and Centaur Chiron show a negative polarization
surge. The Centaur Chiron has the deepest polarization minimum
(-1.5 -- 1.4\,\%). The two TNOs show differing polarization curves: for
Ixion, the negative polarization increases rapidly with phase; for
Quaoar, the polarization is relatively small ($\simeq -0.6$\,\%), and
nearly constant at the observed phase angles. For all three objects,
modelling results suggest that the surface contains an areal mixture
of at least two components with different single-scatterer albedos and
photon mean-free paths.
}
{}

\titlerunning{Polarimetry of Kuiper belt TNO objects and Centaurs} 
\authorrunning{S.~Bagnulo et al.\ }

\maketitle

\section{Introduction}\label{Sect_Introduction}
Transneptunian objects (TNOs) in the Kuiper Belt are considered to
represent one of the oldest and possibly most original population of
solar system bodies that can be observed from Earth. Centaurs are
escapees from the Kuiper Belt through gravitational interaction with
Neptune and the other giant planets. They may eventually become members
of the Jupiter family of comets, or may be ejected from the planet
region due to close encounters with the giant planets.

The intense study of physical properties of TNOs and Centaurs was
triggered by the advent of large telescopes on the ground: besides a
large set of photometric colours, also visible and near-IR spectra of
a number of objects are available now. Polarimetric observations are
more scarce: except for Pluto/Charon system (that was observed
unresolved, e.g., by Kelsey \& Fix \cite{KelFix73}), it was only
recently that broadband polarized radiation of a TNO, the Plutino
28978\,Ixion, has been observed and modelled (Boehnhardt et
al.~\cite{Boeetal04}).

Polarimetry is a powerful tool to investigate the physical properties
of atmosphereless bodies. At small ($\le 30\degr$) phase angles (the
phase angle is the angle between the Sun and the observer as seen from
the object), these objects exhibit a phenomenon of \textit{negative
polarization}: the observed flux perpendicular to the plane
Sun-Object-Observer (the scattering plane) minus the observed flux
perpendicular to that plane, divided by the sum of the two fluxes,
turns to be a negative quantity. This phenomenon, first discovered
through lunar observations by Lyot (\cite{Lyot29}), escapes from
common sense interpretation, since elementary physics tells that
reflected electric vector $\vec{E}$ oscillates predominantly in the
plane perpendicular to the scattering plane rather than in the
scattering plane.  Solar-system objects show two types of angular
dependence of negative polarization: either a smooth phase-angle
change that has the minimum at $\sim 10\degr$ (S-, C- asteroids, Moon)
or a sharp surge with the minimum at $\sim 1-2\degr$ (Saturn rings,
Europa, E-asteroids) (see, e.g., Rosenbush et
al.~\cite{Rosetal02}). Both types of negative polarization, which also
were observed in powdered laboratory samples, are currently
interpreted in terms of enhanced backscattering of multiply scattered
rays (Shkuratov \cite{Shkuratov89}; Muinonen \cite{Muinonen90}).

Observations of negative polarization and simultaneous photometry of
main-belt asteroids and other solar system bodies (see, e.g., Belskaya et
al. \cite{Beletal05}; Rosenbush et al. \cite{Rosetal05}) can be
modelled to infer the properties of the surface texture of these
objects. Faintness of the targets was the main obstacle hampering the
same kind of study in TNOs and Centaurs\footnote{Another difference in
the observing and modelling techniques is that, due to the larger
distance, the observed phase angle range is much smaller for TNOs and
Centaurs than for main-belt asteroids}. Thanks to the advent of the large
telescopes and instruments equipped with polarimetric capabilities,
observations of TNOs and Centaurs are nowadays possible with signal to
noise ratio comparable to that commonly reached for main-belt asteroid
observations with small and middle-size telescopes.

After our first polarimetric study of 28978\,Ixion (Boehnhardt et
al. \cite{Boeetal04}), in this paper we present new polarimetric and
photometric measurements obtained with FORS1 at the ESO Very Large
Telescope (VLT) for a TNO, 50000\,Quaoar, and a Centaur,
2060\,Chiron. We also present a revised modelling of the observed
polarization and photometry of 28978\,Ixion based on a determination
of the geometric albedo that has been recently obtained, and that was
not available at the time of our first modelling effort.

\section{Target summary}\label{Sect_Targets}
Criteria used for target selection are that the targets are bright
enough to allow us to measure the polarization with an error bar
smaller than 0.05\,\% in less than two hours telescope time. With
FORS1 at the ESO VLT, this sets the $R$ magnitude limit to about
20. Another constraint is the possibility to observe the largest
possible phase angle range. Complementary information on the geometric
albedo and surface composition is essential for the modelling
part. Moreover, it is desirable to study members of the various
dynamical and taxonomic groups identified among the TNO population
(Plutinos, Centaurs, classical and scattered-disk Objects). We finally
selected three objects with known physical parameters
(geometric albedos, colours, spectral slopes, and surface composition):
Ixion, Quaoar, and Chiron.

\subsection{28978\,Ixion}
28978\,Ixion, discovered in 2001, belongs to the dynamical class of
Plutinos, and it is one of the largest known TNOs (400--550\,km
according to Stansberry et al. \cite{Stansbe05}).  The visible
spectrum by Marchi et al. (\cite{Maretal03}) is featureless with a
gradient $S'$ of 19.8\,\%/100\,nm. Optical and near-IR (Licandro et
al. \cite{Licetal02}) spectra have been interpreted by Boehnhardt et
al. (\cite{Boeetal04}) using an areal mixture of Titan tholin,
amorphous carbon, water ice, and ice tholin. The same authors also
present a surface model of Ixion based upon their $R$ filter
polarimetry and simultaneous $R$ band photometry of the objects
spanning the phase angle range 0.25\degr\ -- 1.34\degr. In that work,
an $R$-band geometric albedo of 0.1 was assumed for the
modelling. Here we repeat the analysis for the higher $R$-band
geometric albedo now available from Spitzer observations (0.23;
Stansberry, priv.\ comm.).
Further details about the properties of this object are given by
Boehnhardt et al.~(\cite{Boeetal04}).

\subsection{50000\,Quaoar}
50000\,Quaoar is a classical disk object in the Kuiper Belt.  Orbital
elements and red visible colours (Fornasier et al.~\cite{Foretal04})
suggest that the object could be a member of the ``dynamically hot''
population that is supposed to have migrated to the classical disk
only after formation closer to the Sun (Gomes \cite{Gomes03}).  Apart
from Pluto/Charon, 50000\,Quaoar is the only TNO so far for that
disk-resolved photometry could be performed: HST measurements
allowed to determine the overall size and geometric albedo of the
object to be $1260 \pm 190$\,km, and about 0.1, respectively (Brown \&
Trujillo \cite{BroTru04}). The photometric lightcurve of the object
seems to be double-peaked with a period of about 17.6\,h and an
amplitude of 0.13\,mag suggesting an aspherical shape of the body
and/or geometric albedo variations of the surface (Ortiz et 
al.\ \cite{Ortetal03}).

Quaoar visible spectra were obtained by Marchi et al.\
(\cite{Maretal03}) and by Fornasier et al.\ (\cite{Foretal04}). The
reflectivity gradients $S'$ obtained in the two papers are not fully
consistent, and their mean value is $27.6 \pm
0.3$\,\%/100\,nm. Visible spectrum appears to be featureless.

Quaoar has been observed also in the near-infrared by Jewitt and Luu
(\cite{JewLuu04}) at the Subaru 8\,m telescope. The complete spectrum
shows a positive-slope continuum from 0.4 up to 1.3\,$\mu$m, that is
considered typical for the presence of organic materials on its
surface. The spectrum shows strong absorption bands at 1.5 and
2.0\,$\mu$m due to H$_2$O ice with the band at 1.65\,$\mu$m typical
for the crystalline structure in the ice.  A small presence of ammonia
hydrate has also been supposed on the basis of the presence of faint
features at 2.2\,$\mu$m (detected also by Pinilla-Alonso et al.\
\cite{Pinetal04}). This was the first time that the quality of a TNO
spectrum was good enough to distinguish between crystalline and
amorphous ice. The detection of crystalline ice indicates that the
temperature has reached at least 110\,K (critical temperature
necessary for crystallization). This object is large enough to be
cryovolcanically active, and crystalline ice and ammonia hydrate might
be products of this type of activity. Jewitt and Luu
(\cite{JewLuu04}) suggested that Quaoar has been recently resurfaced
either by impacts or by cryovolcanic outgassing or by a combination
of these two processes.

\subsection{2060\,Chiron}
2060\,Chiron is the first discovered (1977) and best observed
Centaur. The intermediate character of the object between TNOs and
comets is apparent from photometric observations that show recurrent
episodes of coma activity (gas and dust) and of stellar appearance
(see for example Meech \& Belton \cite{MeeBel90}, Luu \& Jewitt
\cite{LuuJew90}, Bus et al.\ \cite{Busetal91}, Duffard et al.\
\cite{Dufetal02}).

Chiron was observed spectroscopically in the visible region by many
authors (see Barucci et al.\ \cite{Baretal03}) showing a flat spectrum
with no absorption features. The reflectivity gradient $S'$, which
ranges between $-0.2$ up to $2.3 \pm 0.1$\,\%/100\,nm, seems more
similar to that of C-type asteroids rather than to the mean
reflectance slope of cometary nuclei.  The small variation on the
optical reflectivity gradient could be due to dust production
variation connected to episodes of recurrent cometary activity.

Several spectra obtained in the near-infrared did not show any
features. Only Foster et al.~(\cite{Fosetal99}) and Luu et
al.~(\cite{Luuetal00}) detected a 2\,$\mu$m absorption band suggesting
the presence of H$_2$O ice on Chiron's surface. Later on, Romon-Martin
et al. (\cite{Rometal03}) observed Chiron again during high activity
in the visible and NIR showing a flat behaviour without any spectral
features finding that is compatible with the hypothesis made by Luu et
al.\ (\cite{Luuetal00}) that the detection of water ice in Chiron
spectra would be correlated with its cometary activity level. Such
activity could cause a rain of cometary debris on its surface changing
the surface mantle.  The ice present on the surface, is probably mixed
with dark impurities which mask the spectral bands.

Nucleus properties using multi-wavelength information has revealed an
about 70\,km nucleus of relatively bright geometric albedo (0.17) and
moderate axis ratio (1.16) or surface albedo variations (Groussin et
al.\ \cite{Groetal04}).

From the large number of publications on Chiron (more than 150
to-date) several interesting properties of 2060\,Chiron have been
worked out. However, a synoptic picture of the nucleus and its surface
properties has not yet evolved.

\section{New observations with FORS1}\label{Sect_New_Observations}
Observations of Chiron and Quaoar have been obtained at the ESO VLT
with the FORS1 instrument in service mode during the observing period
from April to September 2004. Until June 1, 2004, FORS1 was attached
at the VLT Unit Telescope 1 (Antu). After that date, FORS1 was moved
to the VLT Unit Telescope 2 (Kueyen).

FORS1 is a multi-mode instrument for imaging and (multi-object)
spectroscopy equipped with polarimetric optics. For the present study,
FORS1 has been used to measure the broadband polarization of Chiron at
six different epochs in the Bessel $BVR$ filters, and to measure the
broadband polarization of Quaoar at five different epochs in the
Bessel $R$ filter (Sect.~\ref{Sect_Polarimetry}). In fact, one series
of Chiron measurements was started on night 2004-08-05/06 and aborted
after the observations in the $R$ filter because the seeing conditions
were too good ($\la 0.6''$) with consequent risk of CCD saturation. 
The series was repeated the following night. From target acquisition 
images, a by-product of polarimetric observations, we could also 
obtain photometry in the Bessel $R$ filter (Sect.~\ref{Sect_Photometry}).
Differential tracking was used for all our observations, so that
long exposure time images of Chiron, obtained in polarimetric mode, could
be combined altogether to search for coma activity
(Sect.~\ref{Sect_Coma_Searching}).

Taking advantage of the flexibility offered by the `VLT service
observing mode', we distributed the observations along a few months as
to obtain data points approximately equally spread over the phase
angle ranges of the targets. We set precise time intervals for the
execution of the observations. In presence of the Moon, the
sky-background is highly polarized, hence we generally tried to avoid
observations with the target close to the Moon, and with a too large
fraction of lunar illumination.  However, \textit{a posteriori} we
found that the presence of the Moon did not jeopardize our
observations. The log of the observations can be inferred from Tables
\ref{Tab_Pol_Chi} to \ref{Tab_Pho_Qua}.

\subsection{Polarimetry}\label{Sect_Polarimetry}
To perform linear polarization measurements, a $\lambda/2$ retarder
waveplate and a Wollaston prism are inserted in the FORS1 optical path
(see Appenzeller \cite{App67}). The $\lambda/2$ retarder waveplate can
be rotated in 22.5\degr\ steps. Stokes~$Q$ and $U$ parameters (defined
as in Shurcliff \cite{Shu62}) are measured by combining the photon
counts (background subtracted) of ordinary and extra-ordinary beams
(\fo\ and \fe, respectively) observed at various retarder waveplate
positions $\alpha$, where $\alpha$ indicates the angle between the
acceptance axis of the ordinary beam of the Wollaston prism and the
fast axis of the retarder waveplate.

In the following, we will always work with the ratios $Q/I$ and $U/I$,
and will adopt the notation:
\[
\pq = \frac{Q}{I}\ \ {\rm and}\ \ \pu = \frac{U}{I}
\]
In the ideal case, \pq\ is
obtained measuring the quantity
\[
r = (-1)^k\ \frac{\fo - \fe}{\fo +\fe}
\]
at any retarder waveplate position $\alpha=k\,45\degr$, 
and \pu\ is obtained measuring the
ratio $r$ at any position $\alpha=k\,45\degr + 22.5\degr$ 
($k=0,\,1,\,2,\,\ldots,\,7$).  The validity of
this assertion can be verified e.g.\ with the help of Eq.~(1.33) of
Landi Degl'Innocenti \& Landolfi (\cite{LanLan04}). In practice, there
are several deviations from the ideal case. For instance, the actual
retardance value of the retarder waveplate may deviate from the
nominal $\pi$ value; the transmission of the ordinary and
extraordinary beam are not identical, even after flat fielding
correction. The effect of these (and other) sources of
\textit{instrumental polarization} can be largely reduced at the first
order by measuring
\begin{equation}
\begin{array}{rcl}
\pq^{ij} &=&
\frac{1}{2}  \Bigg\{
\left(\frac{\fo - \fe}{\fo + \fe}\right)_{\alpha = 2(i-1) \times 45^\circ} -
\left(\frac{\fo - \fe}{\fo + \fe}\right)_{\alpha = (2j-1) \times 45^\circ}
\Bigg\} \\
\pu^{ij} &=&
\frac{1}{2}  \Bigg\{
\left(\frac{\fo - \fe}{\fo + \fe}\right)_{\alpha = 2(i-1) \times 45^\circ + 22.5^\circ} -
\left(\frac{\fo - \fe}{\fo + \fe}\right)_{\alpha = (2j-1) \times 45^\circ + 22.5^\circ}
\Bigg\} \\
\end{array}
\label{Eq_Stokes_QU}
\end{equation}
where $i$ and $j$ are integers numbers\footnote{We note that in a
similar formula that we reported in Boehnhardt et al.\ (\cite{Boeetal04}),
the factor 2 in the indices that denote the angles is missing because of
a typo.}. In the simplest case, linear polarization can be measured from
the observations obtained at four angles of the retarder waveplate:
\[
\begin{array}{rcl}
\pq & = & 
\frac{1}{2}  \Bigg\{
\left(\frac{\fo - \fe}{\fo + \fe}\right)_{\alpha= 0\degr} -
\left(\frac{\fo - \fe}{\fo + \fe}\right)_{\alpha=45\degr}
\Bigg\} \\
\pu & = & 
\frac{1}{2}  \Bigg\{
\left(\frac{\fo - \fe}{\fo + \fe}\right)_{\alpha=22.5\degr} -
\left(\frac{\fo - \fe}{\fo + \fe}\right)_{\alpha=67.5\degr}
\Bigg\} \\
\end{array}
\]

It is convenient (and recommended in the FORS1/2 user manual) to
obtain Stokes $Q$ and $U$ adding up observations obtained with
the retarder waveplate at various positions:
\begin{equation}
\begin{array}{rcl}
\pq&=&\frac{1}{m}\, \sum_{l=1}^{m} \pq^{ll}   \\ [3mm]
\pu&=&\frac{1}{m}\, \sum_{l=1}^{m} \pu^{ll} \;,\\ 
\end{array}
\label{Eq_Sumpol}
\end{equation}
where $m$ represents the number of pairs of observations for each
Stokes parameter, and $\pq^{ll}$ and $\pu^{ll}$ are obtained from
Eq.~(\ref{Eq_Stokes_QU}) setting $i=j=l$. We performed simple
numerical simulations to study the impact on the precision of the
polarimetric measurements of a deviation of the waveplate retardance
from its nominal value (180\degr). We found that using
Eq.~(\ref{Eq_Sumpol}) with $m=2$, a deviation from the nominal value
of the waveplate retardance as large as 5\degr, for the polarization
value observed in our targets, would introduce a spurious contribution
$\ll 0.01$\,\%. Figure~4.1 of FORS1/2 user manual shows
that actual deviation of the retarder waveplate from the nominal value
is well within 5\degr. We conclude that in our data, the effect of
instrumental polarization due to the chromathism of the retarder
waveplate is negligible.

It should be noted that FORS1 is affected by a problem with
spurious linear instrumental polarization that cannot be eliminated
even by using the reduction technique explained above. This spurious
polarization, due to the presence of rather curved lenses in the
collimator, combined with the non complete removal of reflections by
the coatings, is axially symmetric, and smoothly increases from less
than 0.03\,\% on the optical axis to 0.7\,\% at an axis distance of 3
arcmin (in the $V$ band). This problem has been discovered and
investigated by Patat \& Romaniello (\cite{PatRom05}). In our case,
since our targets are always in the center of the field of view, the
problem of the instrumental polarization can be safely ignored.

The error-bar due to photon-noise on the \pq\ or \pu\ measured from a pair of
observations is 
\begin{equation}
\begin{array}{rcl}
\sigma^2_{X^{ij}} & = &
  \left(\left(\frac{\fe}{(\Den)^2}\right)^2 \sigma^2_{\fo} +
  \left(\frac{\fo}{(\Den)^2}\right)^2 \sigma^2_{\fe}\right)_{\alpha=2(i-1) \times 45^\circ + \phi_0 } + \\
             &   &
  \left(\left(\frac{\fe}{(\Den)^2}\right)^2 \sigma^2_{\fo} +
  \left(\frac{\fo}{(\Den)^2}\right)^2 \sigma^2_{\fe}\right)_{\alpha=(2j-1) \times 45^\circ + \phi_0} \;, \\
\end{array}
\label{Eq_Sigma_QU}
\end{equation}
where $\phi_0 = 0$ in case $X$ represents \pq\ and $\phi_0=22.5\degr$
in case $X$ represents \pu. 
When \pq\ and \pu\ are obtained adding up $m$ pairs as in 
Eq.~(\ref{Eq_Sumpol}), the error is given by
\[
\sigma^2_{X} = \frac{1}{m} \sum_{l=1}^{m} \sigma^2_{X^{ll}}
\]
For the sake of simplicity, assuming $\fo
= \fe = \cal{N}$ and also assuming $\cal{N}$ to be the same for all
positions of the retarder waveplate, we obtain
\begin{equation}
\sigma_{X} = \frac{1}{2} \frac{1}{\sqrt{m\cal{N}}}
\label{Eq_Photon_Noise}
\end{equation}
where $\sqrt{\cal{N}}$ represents the signal to noise ratio (SNR)
measured in the individual beam for each of the $m$ exposures (in
other words, $\sqrt{m\cal{N}}$ is the cumulative SNR in each beam).
It appears for instance that to get an error bar on the Stokes
parameter \pq\ or \pu\ of about 0.1\,\%, one should take a pair of
exposures with SNR of about 500 each (integrated over the
individual point-spread function area of each beam of each exposure).

When multiple pairs of exposures are taken, it is useful to study the
distribution of the $\pq^{ij}$ ($\pu^{ij}$) values obtained
substituting $i,j = 1, 2, \ldots m$. In particular, a
$\sigma$-clipping algorithm can be applied to the $\pq^{ij}$ and
$\pu^{ij}$ distributions in order to ``clean'' the data, by rejecting
those values that deviate more than a certain distance from the median
(see Boehnhardt et al.\ \cite{Boeetal04}, and Bagnulo et al.\
\cite{Bagetal05}).  

Stokes~$Q$ and $U$ are usually measured with the instrument position
angle = 0\degr, i.e. to have the acceptance axis of the
ordinary beam of the Wollaston prism aligned to the North Celestial
Meridian (and the acceptance axis of the extra-ordinary beam perpendicular
to it). We then transform the Stokes parameters according to
\begin{equation}
\begin{array}{rcl}
\pq' &=& \phantom{-}\cos(2(\Phi+\pi/2))\, \pq + \sin(2(\Phi+\pi/2))\, \pu\\
\pu' &=&          - \sin(2(\Phi+\pi/2))\, \pq + \cos(2(\Phi+\pi/2))\, \pu\\
\end{array}
\label{Eq_QU_Transform}
\end{equation}
where $\Phi$ is the angle between the direction Object-North Pole
and the direction Object-Sun. This angle can be calculated applying
the four parts formula to the spherical triangle defined by the object
(with coordinates \alphaobj,\deltaobj), the Sun (with coordinates
(\alphasun,\deltasun) and the North celestial pole:
\[
\sin \deltaobj \cos(\alphasun - \alphaobj) = 
\cos(\deltaobj) \tan(\deltasun) - \sin(\alphasun - \alphaobj) \frac{1}{\tan(\Phi)}\;.
\]
This way $\pq'$ \textit{represents the flux perpendicular to the
plane Sun-Object-Earth (the scattering plane) minus the flux parallel
to that plane, divided by the sum of the two fluxes}. 
The angle of maximum polarization is obtained as 
\begin{equation}
\Thetar = \frac{1}{2} \arctan \left(\frac{\pu'}{\pq'}\right) + \Theta_0'
\end{equation}
where
\[
\Theta_0' = \cases {0     &{\rm if} $\pq' > 0$ and $\pu'\ge0$ \cr
                   \pi   &{\rm if} $\pq' > 0$ and $\pu'<0$   \cr
                   \pi/2 &{\rm if} $\pq' < 0$             \cr } \;.
\]
or
\[
\Thetar = \cases{\pi/4 &{\rm if} $\pq' = 0$ and $\pu' > 0$ \cr
                3\pi/4 &{\rm if} $\pq' = 0$ and $\pu' < 0$ \cr
               }
\]
(see Landi Degl'Innocenti \& Landolfi \cite{LanLan04}.)  Incidentally,
it should be noted that the $\Theta_0'$ term is occasionally incorrectly
neglected.

Observations of Chiron were performed with the retarder waveplate at
all positions between 0 and 157.5\degr\ (at 22.5\degr\ steps), i.e.,
setting $m=2$ in Eq.~(\ref{Eq_Sumpol}), using the broadband Bessel
$B$, $V$, and $R$ filters. For Quaoar we used all positions of the
retarder waveplate from 0\degr\ to 337.5\degr, i.e., setting $m=4$ in
Eq.~(\ref{Eq_Sumpol}), using the Bessel $R$ filter.  For each frame,
exposure times $t$ were as follows. For Chiron we set $t=460, 140$,
and 110\,s in the $B$, $V$, and $R$ filter, respectively.  For Quaoar
($R$ filter only), we set $t = 250$\,s. Note that, for each Stokes
parameter we obtained 2 pairs of exposures for Chiron, and 4 pairs of
exposures for Quaoar. Therefore the total shutter time for Chiron
observations was of about 60, 20, and 15\,min in the $B$, $V$, and $R$
filter, respectively, and of about 67\,min for Quaoar ($R$ filter only).
For each observation of Chiron, the typical SNR accumulated in each
beam was about 1700, 1600, and 1600, in the $B$, $V$, and $R$ filters
respectively. For each observation of Quaoar, in the $R$ filter, we
obtained a SNR accumulated in each beam of about 1400.

The photon counts \fo\ and \fe\ were measured via simple aperture
photometry performed on the images, obtained after bias subtraction
and flatfield correction (master flat field was obtained from sky
images obtained during twilight with no polarimetric optics in). More
sophisticated methods based on point spread function (PSF)
fitting are difficult to apply because of the star trailing due to
differential tracking on the moving targets.  Sky background was
measured in an annulus with 5\,\arcsec\ inner radius, centered around
the source, of 2\arcsec\ to 6\arcsec\ width.  The errors on \fo\ and
\fe\ were estimated as explained in the Sect.~3.3.5.8 of Davis
(\cite{Dav87}). \fo, \fe, and their error estimates were measured for
aperture values ranging from 0.6\arcsec to 4\arcsec.


\pq\ and \pu\ values were found to be slightly dependent on the
aperture adopted to measure \fo\ and \fe. This effect was more
critical when the target was in a crowded field and/or in presence of
strong background polarization by the Moon. The final value was 
selected as the one for which the error on \pq\ and \pu\ was minimum,
i.e., usually for aperture values between 1.2\arcsec\ and 2.5\,\arcsec.

\begin{table*}
\caption{The observed polarization of 2060\,Chiron. The meaning of
the various columns is given in the text}
\label{Tab_Pol_Chi}
\centering
\begin{tabular}{cccclcccrrr}
\hline \hline
Date                             &
Time (UT)                        &
\multicolumn{1}{c}{Moon}         &
\multicolumn{1}{c}{}             &
\multicolumn{1}{l}{Sky}          &
Phase angle                      &
Filter                           &
\multicolumn{1}{c}{$\pq'$}       &
\multicolumn{1}{c}{$\pu'$}       &      
\multicolumn{1}{c}{$\Thetar$}     \\ 
\multicolumn{1}{c}{(yyyy mm dd)} & 
\multicolumn{1}{c}{(hh:mm)}      &
\multicolumn{1}{c}{dist.}        &
\multicolumn{1}{c}{FLI}          &
\multicolumn{1}{l}{transp.}      &
\multicolumn{1}{c}{(DEG)}        &
                                 &
\multicolumn{1}{c}{(\%)}         &
\multicolumn{1}{c}{(\%)}         & 
\multicolumn{1}{c}{(DEG)}        \\
\hline
2004 05 27 & 06:37 &139\degr&0.3&CLR    & 3.469 &$R$&$-1.14 \pm 0.04$&$ 0.02 \pm 0.04$&$89.4 \pm 1.0$\\
2004 05 27 & 07:03 &        &   &       & 3.468 &$V$&$-1.20 \pm 0.04$&$ 0.13 \pm 0.04$&$86.9 \pm 0.9$\\
2004 05 27 & 07:52 &        &   &       & 3.467 &$B$&$-1.24 \pm 0.04$&$-0.04 \pm 0.05$&$90.9 \pm 1.1$\\[2mm]
                                                                         
2004 06 29 & 03:19 & 64\degr&0.4&CLR?   & 1.410 &$R$&$-1.40 \pm 0.05$&$ 0.05 \pm 0.04$&$89.1 \pm 0.8$\\
2004 06 29 & 03:44 &        &   &       & 1.409 &$V$&$-1.35 \pm 0.05$&$-0.08 \pm 0.04$&$91.6 \pm 0.9$\\
2004 06 29 & 04:33 &        &   &       & 1.407 &$B$&$-1.32 \pm 0.04$&$ 0.08 \pm 0.05$&$88.2 \pm 1.2$\\[2mm]
                                                                         
2004 08 06 & 05:53 & 94\degr&0.7&PHO    & 1.766 &$R$&$-1.40 \pm 0.04$&$ 0.00 \pm 0.04$&$89.9 \pm 0.8$\\[2mm]
                                                                         
2004 08 07 & 03:46 &105\degr&0.7&THN (c)& 1.828 &$R$&$-1.33 \pm 0.04$&$-0.01 \pm 0.04$&$90.2 \pm 0.8$\\
2004 08 07 & 04:13 &        &   &       & 1.829 &$V$&$-1.46 \pm 0.03$&$-0.14 \pm 0.03$&$92.7 \pm 0.7$\\
2004 08 07 & 05:00 &        &   &       & 1.832 &$B$&$-1.52 \pm 0.04$&$-0.09 \pm 0.05$&$91.8 \pm 0.9$\\[2mm]
                                                                         
2004 08 13 & 00:17 &167\degr&0.9&PHO (c)& 2.219 &$R$&$-1.33 \pm 0.04$&$ 0.08 \pm 0.04$&$88.3 \pm 1.0$\\
2004 08 13 & 00:42 &        &   &       & 2.221 &$V$&$-1.34 \pm 0.04$&$-0.08 \pm 0.04$&$91.6 \pm 0.9$\\
2004 08 13 & 01:31 &        &   &       & 2.223 &$B$&$-1.45 \pm 0.03$&$-0.19 \pm 0.05$&$93.8 \pm 1.1$\\[2mm]
                                                                         
2004 09 01 & 00:56 & 74\degr&0.5&THN? (c)& 3.327&$R$&$-1.12 \pm 0.04$&$ 0.05 \pm 0.04$&$88.7 \pm 1.1$\\
2004 09 01 & 01:21 &        &   &       & 3.328 &$V$&$-1.09 \pm 0.05$&$-0.08 \pm 0.04$&$92.1 \pm 1.2$\\
2004 09 01 & 02:10 &        &   &       & 3.330 &$B$&$-1.15 \pm 0.05$&$-0.24 \pm 0.06$&$95.9 \pm 1.6$\\[2mm]
                                                                         
2004 09 27 & 01:24 & 56\degr&0.4&CLR?   & 4.232 &$R$&$-0.89 \pm 0.07$&$ 0.03 \pm 0.05$&$89.1 \pm 1.8$\\
2004 09 27 & 01:49 &        &   &       & 4.232 &$V$&$-1.05 \pm 0.05$&$-0.20 \pm 0.05$&$95.4 \pm 1.7$\\
2004 09 27 & 02:38 &        &   &       & 4.233 &$B$&$-0.87 \pm 0.09$&$-0.10 \pm 0.08$&$93.3 \pm 3.0$\\[2mm]
\hline
\end{tabular}
\end{table*}

\begin{table*}
\caption{The observed polarization of 50000\,Quaoar. The meaning of
the various columns is given in the text}
\label{Tab_Pol_Qua}
\centering
\begin{tabular}{cccclcccrrr}
\hline \hline
Date                             &
Time (UT)                        &
\multicolumn{1}{c}{Moon}         &
\multicolumn{1}{c}{}             &
\multicolumn{1}{l}{Sky}          &
Phase angle                      &
Filter                           &
\multicolumn{1}{c}{$\pq'$}       &
\multicolumn{1}{c}{$\pu'$}       &      
\multicolumn{1}{c}{$\Thetar$}     \\ 
\multicolumn{1}{c}{(yyyy mm dd)} & 
\multicolumn{1}{c}{(hh:mm)}      &
\multicolumn{1}{c}{dist.}        &
\multicolumn{1}{c}{FLI}          &
\multicolumn{1}{l}{transp.}      &
\multicolumn{1}{c}{(DEG)}        &
                                 &
\multicolumn{1}{c}{(\%)}         &
\multicolumn{1}{c}{(\%)}         & 
\multicolumn{1}{c}{(DEG)}        \\
\hline
2004 04 18 & 04:43 & 47\degr&1.0&CLR     & 0.952 &$R$&$-0.64 \pm 0.05$&$ 0.14 \pm 0.08$&$83.9 \pm 3.9$\\
2004 05 13 & 07:38 & 93\degr&0.8&CLR?    & 0.496 &$R$&$-0.50 \pm 0.06$&$-0.14 \pm 0.05$&$97.7 \pm 3.6$\\
2004 05 26 & 03:37 &107\degr&0.2&PHO     & 0.252 &$R$&$-0.49 \pm 0.06$&$ 0.10 \pm 0.05$&$84.3 \pm 3.5$\\
2004 07 10 & 01:44 &118\degr&0.7&PHO (c) & 0.797 &$R$&$-0.53 \pm 0.04$&$ 0.09 \pm 0.05$&$85.3 \pm 3.1$\\
2004 08 11 & 02:34 & 75\degr&0.8&CLR?    & 1.231 &$R$&$-0.65 \pm 0.04$&$-0.01 \pm 0.06$&$90.4 \pm 2.6$\\
\hline
\end{tabular}
\end{table*}

The results of the polarimetric measurements are given in
Tables~\ref{Tab_Pol_Chi} and \ref{Tab_Pol_Qua}, that are organized as
follows. Columns 1 and 2 give the epoch of the observations (date and
UT time).

Columns~3 and 4 give Moon angular distance and fraction of lunar
illumination (FLI), respectively.

Column 5 gives an estimate of the night time sky conditions: THN =
thin cirrus, CLR= clear, PHO = photometric. The classification given
in these Tables is based on our inspection of reduction products and
of the atmospheric monitors of the observatory at \\ {\tt
http://www.eso.org/gen-fac/pubs/astclim/}\\ \ \ {\tt
forecast/meteo/CIRA/images/repository/lossam/} \\ and is somewhat
arbitrary. Those nights when no photometric standard stars were
observed, or when the zeropoints were not considered stable for the
entire night, are indicated with a question mark. Sky transparency
does not affect the precision of the polarization measurements, but it
does affect the precision of the photometry (see
Tables~\ref{Tab_Pho_Chi} and \ref{Tab_Pho_Qua}). Note that the
precision of the measurements (both polarimetric and photometric)
depends also on how the field of view close to the target is crowded
with background objects. This situation of course changes from epoch
to epoch. Observations labelled with (c) are hampered to some extent
by a crowded background. In the case of Chiron observations on
2004-09-01 the situation was probably especially critical.

Column 6 gives the Sun-Target-Observer angle, i.e., the
target's apparent phase angle as seen at observer's location,
expressed in degrees. This phase angle was obtained from the
object ephemeris calculated at the JPL's
solar system dynamics WWW site at {\tt http://ssd.jpl.nasa.gov}.

Column~7 specifies the broadband filter used for the observations.

Columns~8, 9, and 10 give \pq', \pu', and the angle of maximum
polarization $\Thetar$ , respectively, after the transformation of
Eq.~(\ref{Eq_QU_Transform}).  It should be noted that for both Chiron
and Quaoar, the measured $\pu'$ is generally consistent with zero
(equivalent to the fact that $\Thetar$ is generally consistent with
90\degr). This means that the principal axes of the polarization
ellipse are aligned with the coordinate axes perpendicular and
parallel to the scattering plane.  Furthermore, the observed $\pq'$,
i.e., the flux perpendicular to the scattering plane minus the flux
parallel to that plane, divided by the sum of the two fluxes, is
always a \textit{negative} quantity.  This means that the direction of
the polarization is included in the scattering plane (this case is
normally referred to as ``negative polarization''), in contrast to
what is expected for a dielectric medium, i.e., that the direction of
polarization be perpendicular to the scattering plane (this case is
normally referred to as ``positive polarization'').

Polarimetric measurements of Quaoar and Chiron are plotted against
object phase angle in Figs.~\ref{Fig_Pol_Qua} and \ref{Fig_Pol_Chi}.
As far as Chiron observations are concerned, it appears that the
observed polarization does not depend on the filter used, at least
when considering the error bars of the observations. The absolute
value of the polarization increases as the phase angle decreases,
perhaps reaching a minimum around phase 1.5-2.0 deg. Observations of
this object at smaller phase angles were unfortunately not taken
although originally scheduled in our service observing campaign.

\subsection{Photometry}\label{Sect_Photometry}
The object magnitudes were measured in the acquisition images
obtained in the $R$ Bessel filter with no polarimetric optics in the
light path. Exposure times were 5\,s for Chiron, and 30\,s for Quaoar.

Since our program was aimed at polarization measurements, we did not
obtain a number of observations of photometric standard stars
sufficient to estimate precise zeropoints and extinction coefficients
for the various nights. In most of the cases, the zeropoints were
obtained from only one frame obtained during the night when our
observations were executed (and not necessarily close to them in
time). In some cases, no photometric standard stars were observed at
all (see Tables~\ref{Tab_Pol_Chi} and \ref{Tab_Pol_Qua}). In these
cases we adopted the values measured in the nearest night with similar
sky conditions. For the extinction coefficient $K_R$ and the colour
term $k_{VR}$ we used the values estimated for the P73 ESO observing
period published at\\ 
{\tt http://www.eso.org/observing/dfo/quality/FORS1/}\\ \ \ {\tt
qc/photcoeff/photcoeffs\_fors1.html}\\ i.e.,
\[
\begin{array}{rcl}
K_R    &=& 0.045  \pm 0.019 \\ 
k_{VR} &=& 0.0090 \pm 0.0017 \\
\end{array}
\]
For the Quaoar colour indices we adopted $V-R = 0.64 \pm 0.01$ (Tegler
et al. \cite{Tegetal03}), and for Chiron $V-R = 0.37 \pm 0.03$ (Davies
et al. \cite{Davetal98}). Finally, reduced magnitude $R$ of the
target was obtained by taking into account the distance Sun-Object
$r$ and the distance Earth-Object $\Delta$ at the epoch of the
observations, and calculating the reduced magnitude
\begin{equation}
R = m_R -5\,\log (r)- 5\,\log (\Delta) 
\label{Eq_Reduced_Mag}
\end{equation}
where $m_R$ is the observed $R$-band magnitude.

The results of our photometry are given in Tables~\ref{Tab_Pho_Chi}
and \ref{Tab_Pho_Qua}. 

From a linear least-square fit of the data, we found that for Chiron
the $H_R$ absolute magnitude at zero phase angle is $5.52 \pm
0.07$\,mag, and the slope of the opposition surge is $0.045 \pm
0.023$\,mag/deg.  Previous numerous photometric observations of Chiron
(for summary see Groussin et al. \cite{Groetal04}) have shown
considerable variations of absolute magnitude with heliocentric
distance attributed to its sporadic cometary behavior. The absolute
magnitude of Chiron in 2004 is brighter as compared to the
observations in 1988-1990 made at the same heliocentric distances (Bus
et al.\ \cite{Busetal91}). It may indicate that we observed Chiron
closely after an activity period which may have caused partial
resurfacing of the object. This is why we paid a special attention
to search for a coma around Chiron (see
Sect.~\ref{Sect_Coma_Searching}).

For Quaoar, the $H_R$ absolute magnitude at zero phase angle, obtained
from a \textit{linear fit} is $2.16 \pm 0.05$\,mag, and the linear
slope is $0.16 \pm 0.06$\,mag/deg. The determined value of absolute
magnitude coincides with that used for Quaoar for geometric albedo
determination (Brown \& Trujillo \cite{BroTru04}).

Note that $H_R$ absolute magnitudes at zero phase angles and linear
slopes will be re-estimated based on our modelling technique in
Sect.~\ref{Sect_Modelling}.

\begin{table}
\caption{The observed $R$ photometry of 2060\,Chiron}
\label{Tab_Pho_Chi}
\centering
\begin{tabular}{ccccc}
\hline \hline
Date                             &
Time (UT)                        &
Phase angle                      &
\multicolumn{1}{c}{$m_R$}        &
\multicolumn{1}{c}{$R$}         \\
\multicolumn{1}{c}{(yyyy mm dd)} & 
\multicolumn{1}{c}{(hh:mm)}      &
\multicolumn{1}{c}{(DEG)}        &
(mag)                            &
(mag)                           \\
\hline
2004 05 27 & 06:23 & 3.470 & $16.64 \pm 0.04$& 5.68 \\
2004 06 29 & 03:04 & 1.411 & $16.48 \pm 0.05$& 5.56 \\
2004 08 06 & 05:40 & 1.766 & $16.49 \pm 0.04$& 5.54 \\
2004 08 07 & 03:33 & 1.828 & $16.65 \pm 0.04$& 5.69 \\
2004 08 13 & 00:02 & 2.219 & $16.55 \pm 0.05$& 5.58 \\
2004 09 01 & 00:39 & 3.326 & $16.65 \pm 0.04$& 5.63 \\
2004 09 27 & 01:10 & 4.232 & $16.84 \pm 0.04$& 5.74 \\
\hline
\end{tabular}
\end{table}
\begin{table}
\caption{The observed $R$ photometry of 50000\,Quaoar}
\label{Tab_Pho_Qua}
\centering
\begin{tabular}{ccccc}
\hline \hline
Date                             &
Time (UT)                        &
Phase angle                      &
\multicolumn{1}{c}{$m_R$}        &
\multicolumn{1}{c}{$R$}         \\
\multicolumn{1}{c}{(yyyy mm dd)} & 
\multicolumn{1}{c}{(hh:mm)}      &
\multicolumn{1}{c}{(DEG)}        &
(mag)                            &
(mag)                           \\
\hline
2004 04 18 & 03:59 & 0.952 & $18.60 \pm 0.05$& 2.27 \\
2004 05 13 & 06:54 & 0.497 & $18.60 \pm 0.04$& 2.27 \\
2004 05 26 & 02:52 & 0.252 & $18.51 \pm 0.04$& 2.19 \\
2004 07 10 & 00:59 & 0.796 & $18.57 \pm 0.04$& 2.24 \\
2004 08 11 & 01:49 & 1.231 & $18.73 \pm 0.04$& 2.38 \\
\hline
\end{tabular}
\end{table}

\subsection{Search for a coma around 2060\,Chiron}\label{Sect_Coma_Searching}
\begin{figure}
\includegraphics*[height=9cm,angle=270]{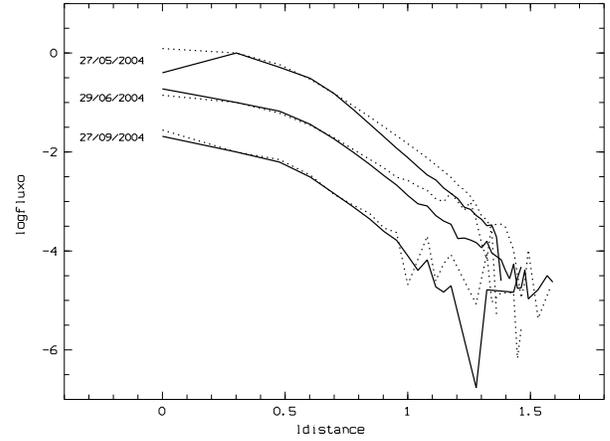}
\caption{
Coma search around 2060 Chiron. Solid lines show the Chiron radial
profiles, the dotted lines show the radial profiles of the comparison
stars. The $x-$axis represents the logarithmic distance in pixels from
the centre of the Chiron (or the stars) PSF images. 
The $y-$axis represents the flux (in arbitrary units).
}
\label{Fig_Coma}
\end{figure}

Visual inspection and measurements of the full-width-at-half-maximum
of the Chiron and neighbouring star images did not reveal any
indications for the presence of a coma around the Centaur. For a more
thorough analysis we calculated the radial flux profile measured in
concentric rings around the object, and compared it with the radial
profile across the star trails, averaged along the trail direction and
for both sides of the trail. The resulting object and star profiles
are normalized to unity brightness at 'one pixel' central distance and
to zero at background distance 30\,pixels from the
center. Figure~\ref{Fig_Coma} shows the results for three observing
dates (for other dates the presence of star trails close to the object
image jeopardized the accuracy of this analysis). Since the radial
profile of Chiron (solid line) is basically identical to that of the
comparison stars (dotted line) or falls at slightly lower flux levels,
we conclude that a coma around Chiron is not present or it is well
beyond our detection limit (order of 31\,mag/arcsec) - if present at
all. Hence, we assume that the polarimetry of the object is not
contaminated by dust around the object.

\section{Modelling of the observed polarization}\label{Sect_Modelling}
\begin{table} 
\caption{ 
The best fit coherent-backscattering model parameters for Ixion,
Quaoar, and Chiron. We give the single-scattering albedos
$\tilde{\omega}$ and dimensionless mean free paths $k\ell$ for the
dark (subscript $d$) and bright components ($b$), the weight of the
dark component $w_d$, the rms values of the polarimetric fits, as well
as the $R$-band absolute magnitudes $H_R$ and slope parameters $k_R$.
}
\label{Table_Results}
\begin{tabular}{llll}
\hline\hline
                   & Ixion       & Quaoar      & Chiron       \\ 
\hline
$\tilde{\omega}_d$ &   0.45      &   0.35      &   0.15       \\
$\ell_d$           & 250         & 300         & 120          \\
$\tilde{\omega}_b$ &   0.80      &   0.50      &   0.60       \\
$\ell_b$           &  20         &  10         & 500          \\
$w_d$              &   0.74      &   0.46      &   0.14       \\
rms                &   0.029\,\% &   0.069\,\% &   0.067\,\%  \\
$H_R$              &   3.25      &   2.15      &   5.41       \\
$k_R$              &   0.12\,deg$^{-1}$&   0.11\,deg$^{-1}$&   0.041\,deg$^{-1}$\\
\hline
\end{tabular}
\end{table}
\begin{figure}
\includegraphics*[height=11cm]{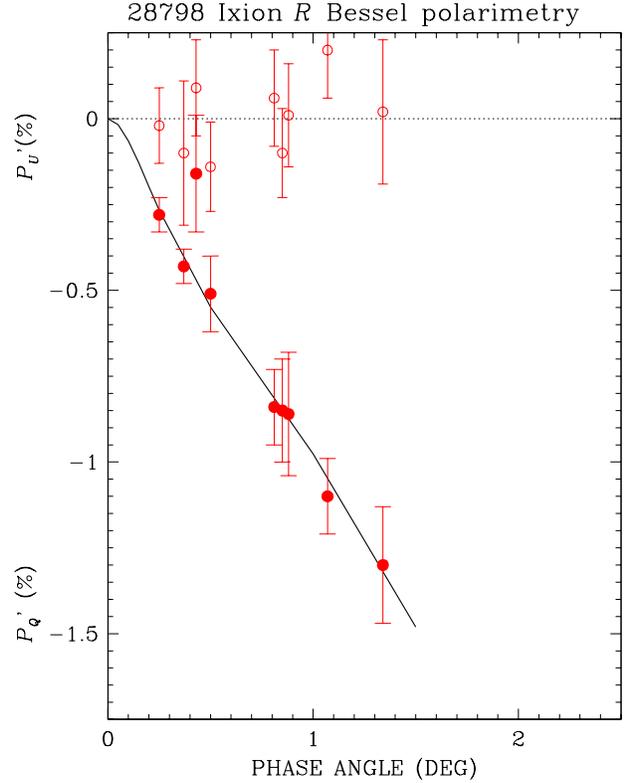}
\caption{
The observed broadband polarization of 28978\,Ixion.  The solid
symbols refer to $\pq'$, i.e., the flux perpendicular to the plane
Sun-Object-Earth minus the flux parallel to that plane, divided by the
sum of the two fluxes. The empty circles refer to $\pu'$ (measured in
the scattering plane as explained in the text). Solid line represents
the modelling fit to $\pq'$.  The dotted line shows the expected
$\pu'$ null values.
} 
\label{Fig_Pol_Ixi}
\end{figure}
\begin{figure}
\includegraphics*[height=11cm]{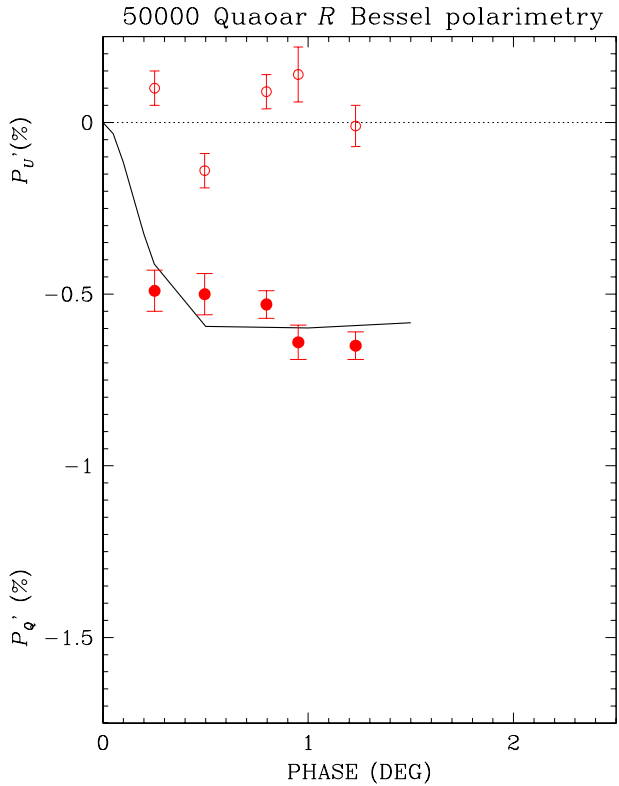}
\caption{ 
The broadband polarization of 50000\,Quaoar observed in the Bessel $R$
filter. As in Fig.~\ref{Fig_Pol_Ixi}, solid symbols refer to $\pq'$,
and open symbols refer to $\pu'$. Solid line represents the modelling fit 
to $\pq'$. Dotted lines show the expected null $\pu'$ values. 
}
\label{Fig_Pol_Qua}
\end{figure}
\begin{figure}
\includegraphics*[height=11cm]{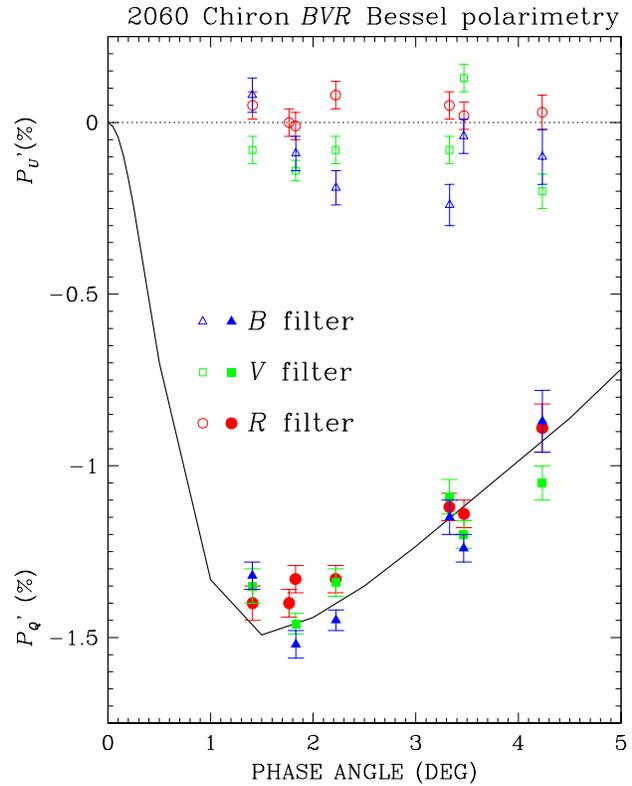}
\caption{ 
The observed broadband polarization of 2060\,Chiron. As in
Fig.~\ref{Fig_Pol_Ixi}, the solid symbols refer to $\pq'$, and the
empty symbols refer to $\pu'$.  Triangles, squares, and circles refer
to the polarization measured in the Bessel $B$, $V$, and $R$ filters,
respectively. Solid line represents the modelling fit to $\pq'$ for
the $R$ filter. The dotted line shows the expected $\pu'$ null values.
} 
\label{Fig_Pol_Chi}
\end{figure}
\begin{figure}
\includegraphics*[height=11cm]{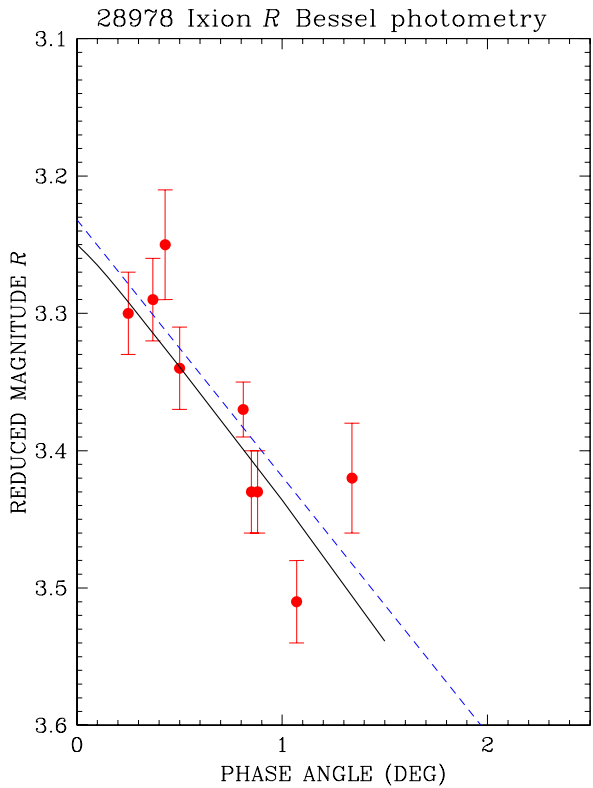}
\caption{ 
The observed reduced magnitude $R$ of Ixion. The dashed line
represents the best fit obtained with a straight line. The solid line
represents the model fit. The scattering around the fits is fully
within the amplitude of intrinsic variability of the object.
}
\label{Fig_Pho_Ixi}
\end{figure}
\begin{figure}
\includegraphics*[height=11cm]{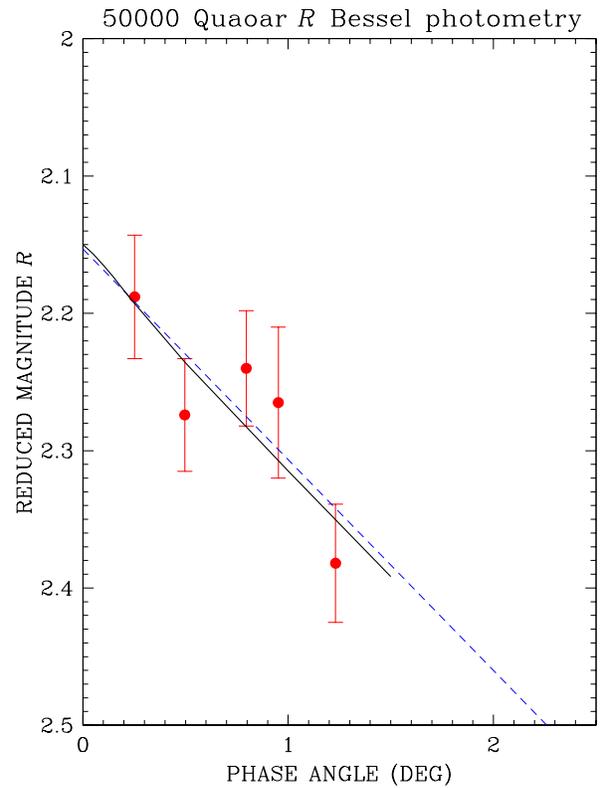}
\caption{
The observed reduced magnitude $R$ of Quaoar.
Symbols are the same as Fig.~\ref{Fig_Pho_Ixi}.
}
\label{Fig_Pho_Qua}
\end{figure}
\begin{figure}
\includegraphics*[height=11cm]{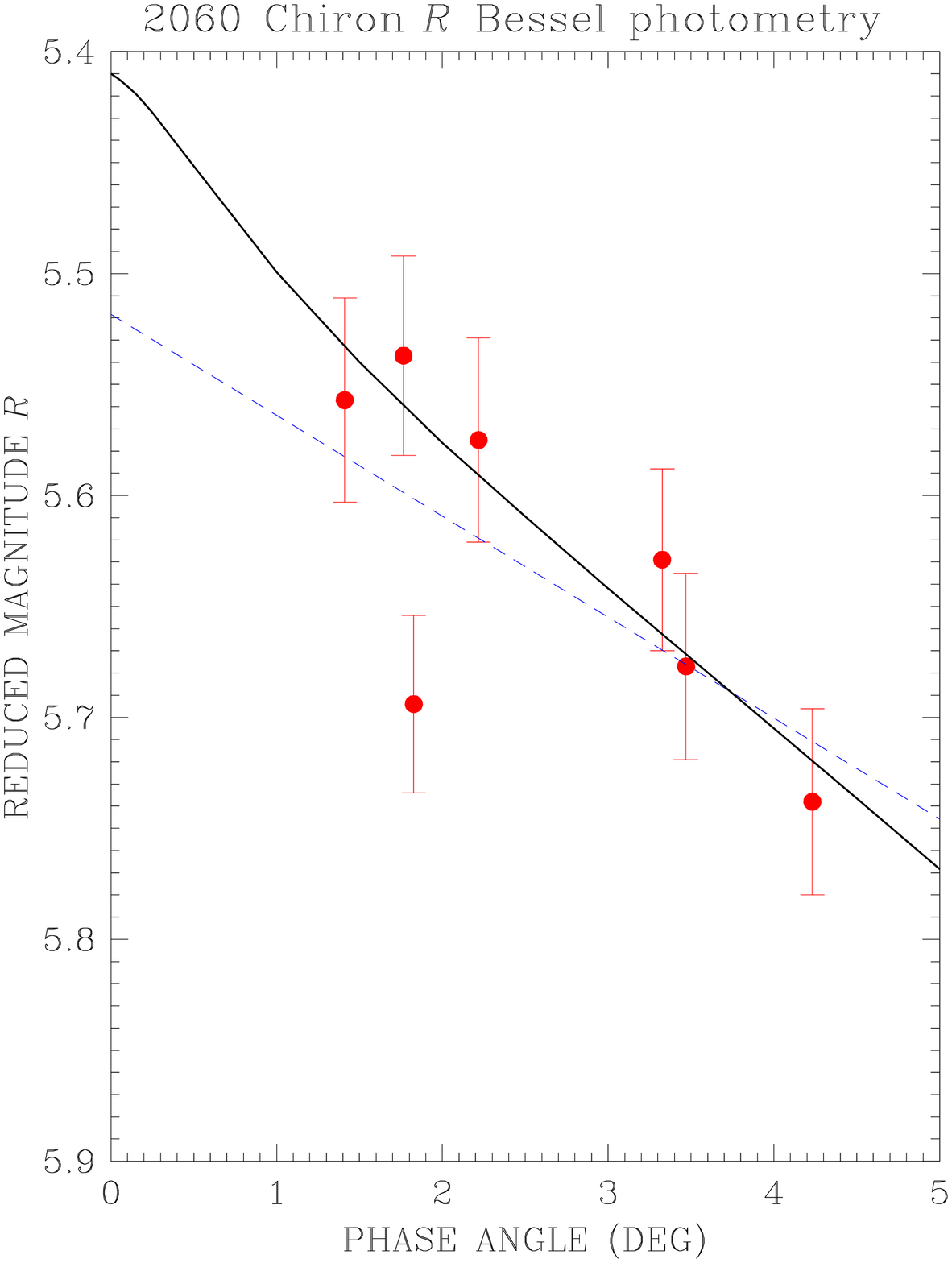}
\caption{
The observed reduced magnitude $R$ of Chiron.
Symbols are the same as Fig.~\ref{Fig_Pho_Ixi}.
}
\label{Fig_Pho_Chi}
\end{figure}
We interpret the polarimetric and photometric phase curves of Ixion,
Quaoar, and Chiron through extensive numerical simulations of coherent
backscattering by Rayleigh scatterers. Note that we avoid making
an assumption that the fundamental scatterers responsible for the
coherent backscattering contribution would be the single particles in
the regolith. With the phenomenological modeling currently including
the first multipole contribution of an electric dipole, we allow for
the possibility that the fundamental scatterers can be the volume and
surface inhomogeneities within and on the particles, respectively.
Shadowing among the regolith particles is known to contribute to the
opposition effect but not to the negative polarization surge. Here, as
in Boehnhardt et al. (\cite{Boeetal04}), the shadowing effect
manifests itself in the residual slope parameter resulting from the
combined coherent backscattering modeling on the polarimetry and
photometry.

The coherent-backscattering mechanism is a multiple-scattering
mechanism for scattering orders higher than the first one. For a
recent review of the coherent-backscattering mechanism and its
relevance in asteroid studies, see Muinonen et al. (\cite{Muietal02})
and Muinonen (\cite{Muinonen04}). Note that the computational
technique for coherent backscattering accounts for all orders of
scattering contributing in a non-negligible way to the backscattering
peaks and polarization surges.

In the following we limit ourselves to illustrate the mechanism at the
second order of scattering. Let us consider a semi-infinite random
medium of discrete scatterers, constrained by a plane-parallel
boundary with free space, and an electromagnetic plane wave incident
on the random medium from the free space. Let us assume that the
incident wave interacts with one of the scatterers, giving rise to
first-order scattering, and that, subsequently, the first-order
scattered field interacts with another scatterer, giving rise to
second-order scattering. Let us assume that the field scattered by the
second scatterer escapes the medium and is detected by the observer in
the free space.  Now there is the reciprocal sequence of scatterings
in the opposite direction involving the very same two scatterers. In
the exact backward scattering direction, due to the identical lengths
of the propagation paths, the two reciprocal wave components interfere
constructively whereas, in other directions, the interference
characteristics vary. After configurational averaging, a
backscattering peak results in the proximity of the backward
scattering direction. Focusing in on the polarization characteristics,
the interference is selective and tends to invert the linear
polarization characteristics in single scattering: for positively
polarizing single scattering, coherent backscattering tends to result
in negative polarization. The illustration can be readily
generalized to higher orders of scattering. In the modeling that
follows, all relevant orders of scattering are taken into account.

For the modelling of Ixion, Quaoar, and Chiron, we carried out
coherent backscattering computations for a total of 360 spherical
media of Rayleigh scatterers, as follows: we assumed 18 different
single-scattering albedos
\[
\tilde{\omega}=0.05, 0.10, \ldots, 0.90
\]
and 20 different dimensionless mean free paths 
\[
\begin{array}{rcl}
k\ell = 2\pi \ell/\lambda &=& 10, 20, 30,\ \ldots,\ 100, 120, 140,\ \ldots, \\
                          & & 200, 250, 300,\ \ldots,\ 400, 500 \\
\end{array}
\]
where $k$ and $\lambda$ are the wave number and wavelength, respectively.

As to the $R$-band geometric albedos $p_R$, we adopted $p_R=0.23$ for
Ixion (Stansberry, priv.\ comm.), 0.11 for Quaoar
(calculated using the diameter from Brown \& Trujillo \cite{BroTru04}
and our determination of the absolute magnitude), and 0.17 for Chiron
(Barucci et al. \cite{Baretal04}). In our previous interpretation of
Ixion polarimetric data (Boehnhardt et al. \cite{Boeetal04}), we had
assumed $p_R = 0.1$.

We compared the polarimetric observations against the spherical media
composed of monodisperse Rayleigh scatterers with given
single-scattering albedo and mean free path (two parameters). The fits
were poor: for a fixed geometric albedo, the monodisperse
Rayleigh-scattering model tends to result in polarizations that are
substantially pronounced as compared to the polarizations observed.

As in Boehnhardt et al. (\cite{Boeetal04}), we then studied a
two-component Rayleigh-scattering model consisting of dark and bright
scatterers. There are five parameters in such a model: two
single-scattering albedos and two mean free paths, and the weight
factor for the dark component (one minus the weight factor of the
bright component). Fixing the geometric albedo fixes the weight
factor, reducing the number of free parameters to four. After a
systematic study of physically realistic combinations of the two kinds
of scatterers, satisfactory fits were obtained for the polarizations
of all three objects. For Ixion, the data point at phase angle
0.43\degr\ was omitted as an outlier. The model parameters and rms
values of the fits are summarized in Table~\ref{Table_Results} and the
actual fits are depicted in Figs.~\ref{Fig_Pol_Ixi} --
\ref{Fig_Pol_Chi}. After the polarization fits, approximate brightness
fits were obtained by varying the absolute magnitude $H_R$ and slope
parameter $k_R$ of a linear phase dependence multiplying the coherent
backscattering contribution (see Fig.~\ref{Fig_Pho_Ixi} --
\ref{Fig_Pho_Chi}). For Chiron, the observation at 1.8\degr\ phase
angle was omitted as outlier.

For all three objects, the dark component shows a mean free path
substantially longer than the wavelength. For Chiron, the mean free
path of the bright component is also considerably longer than the
wavelength. The dark components of Ixion and Quaoar resemble one
another and the bright component of Chiron: this is concluded from the
similarity of the parameters of the corresponding scatterers on the
surfaces of the three objects. The phase curves of Ixion and Chiron
resemble each other to an extent where their combined polarimetric and
photometric data could be explained using a single two-component
scattering model.

The polarimetric observations suggest that Ixion and Chiron could have
similar surface structure.  It is notable that the geometric albedo
ranges of Ixion and Chiron overlap, whereas Quaoar stands out as a
darker object. More observational and theoretical work is required for
further conclusions.

\section{Discussion and Conclusions}\label{Sect_Discussion}
We have presented the results of the first polarimetric observations
for a Centaur and a classical disk object in the Kuiper Belt. Together
with polarimetric observations of Ixion (Boehnhardt et al.,
\cite{Boeetal04}), a representative of the Plutino population in the
belt, they give us a first idea about the behavior of linear
polarization and intensity of the light reflected by surfaces of
Kuiper belt objects and Centaurs. Two Kuiper Belt objects, Ixion and
Quaoar, were observed practically in the same range of phase angles
(0.25\degr--1.3\degr). They have shown completely different
polarization-phase behaviour. For Ixion the negative polarization
rapidly increases with phase angle ($-1.02 \pm 0.025$\,\%/deg), while
for Quaoar the polarization degree is changed very slowly ($-0.17 \pm
0.10$\,\%/deg).
Our polarimetric observations of Chiron were made for larger phase
angles, 1.4\degr--4.2\degr. They are characterized by a pronounced
branch of negative polarization with a minimum of 1.4--1.5\,\% at
phase angles of 1.5\degr--2.0\degr. Such polarimetric characteristics
are unique among Solar System bodies. For the majority of Solar System
bodies the polarization at the phase angle 1\degr\ is characterized by
values within 0.1--0.5\,\%. The largest values of the polarization at 1\degr\
measured for non-TNO objects were 0.83\,\% for the dark side of
Iapetus (albedo = 0.05) and 0.73\,\% for Saturn Ring A (albedo = 0.75)
(Rosenbush et al.\ \cite{Rosetal02}) that is noticeably smaller than the values
obtained at our observations of Ixion and Chiron. Polarimetric
observations of Chiron at smaller phase angles are urgently needed for
a better modelling, and to compare Chiron data with the polarimetric
curve of Ixion and Quaoar. Further observations would be also desirable
to identify the inversion angle.

Our modelling has shown that the possible way to explain observed
polarization properties of KBOs and Centaur is to assume two-component
surface media consisting of dark and bright scatterers. Such a model
succeeds in fitting all observed polarimetric and photometric
characteristics of the three objects and the two-component modelling
is realistic. A more thorough theoretical study is beyond the scope of
the present article and should be carried out in the nearest future.

\begin{acknowledgements}
The authors wish to thank D.\ Rabinowitz for sharing his unpublished
photometric data of Quaoar which have helped us to verify the
calibration of our data, E.~Landi Degl'Innocenti and M.\ Landolfi for
their help to write Sect.~3, and O.~Hainaut for his help to identify
adequate observing periods for our targets (i.e., with no or little
risk of background star confusion).

\end{acknowledgements}

\end{document}